# Current-driven magnetization switching in CoFeB/MgO/CoFeB magnetic tunnel junctions

Jun HAYAKAWA[1,2], Shoji IKEDA[2], Young Min LEE[2], Ryutaro SASAKI[2], Toshiyasu MEGURO[2], Fumihiro MATSUKURA[2], Hiromasa TAKAHASHI[1,2], and Hideo OHNO[2]

*1. Hitachi, Ltd., Advanced Research Laboratory, 1-280 Higashi-koigakubo, Kokubunji-shi, Tokyo 185-8601, Japan*

*2. Laboratory for Nanoelectronics and Spintronics, Research Institute of Electrical Communication, Tohoku University, 2-1-1 Katahira, Aoba-ku, Sendai 980-8577, Japan*

Current-driven magnetization switching in low-resistance $Co_{40}Fe_{40}B_{20}$/MgO/$Co_{40}Fe_{40}B_{20}$ magnetic tunnel junctions (MTJs) is reported. The critical-current densities $J_c$ required for current-driven switching in samples annealed at 270 $^{\circ}$C and 300 $^{\circ}$C are found to be as low as 7.8 x $10^5$ A/cm$^2$ and 8.8 x $10^5$ A/cm$^2$ with accompanying tunnel magnetoresistance (TMR) ratios of 49% and 73 %, respectively. Further annealing of the samples at 350 $^{\circ}$C increases TMR ratio to 160 %, while accompanying $J_c$ increases to 2.5 x $10^6$ A/cm$^2$. We attribute the low $J_c$ to the high spin-polarization of tunnel current and small $M_sV$ product of the CoFeB single free layer, where $M_s$ is the saturation magnetization and $V$ the volume of the free layer.

Spin-polarized currents exert torque on magnetization that can switch the magnetization direction once the current density becomes sufficiently high[1,2]. This current-driven magnetization switching has been demonstrated on a number of metallic current-perpendicular-to-plane giant magnetoresistance (CPP-GMR) pillars[3-8] which exhibit magnetoresistance ratio of 0.5-5 %. The critical current density ($J_c$) of these devices varies from mid $10^6$ to $10^8$ A/cm$^2$ depending on the employed structure. Current-driven switching is preferred over conventional cross-point current-generated magnetic field writing in magnetic random access memories (MRAMs), because the required absolute current scales with the area of the device. In order for this approach to be viable in the 90 nm technology node and beyond, $J_c$ must be lower than $10^6$ A/cm$^2$ to be driven by a MOS transistor that can deliver typically 100 μA per 100 nm gate width. For MRAM, these devices must exhibit high magnetoresistance ratio at the same time. High tunnel magnetoresistance (TMR) ratio, much higher than those reported for the conventional MTJs that use AlO$_x$ as a barrier material and higher than the earlier studies on MTJs based on Fe/MgO/Fe [9], has recently been realized in single-crystal Fe/MgO/Fe MTJs [10,11], highly oriented (001) CoFe/MgO/CoFe MTJs[12], and sputtered CoFeB/MgO/CoFeB MTJs[13-15], as predicted by theoretical studies.[16-18] It is therefore essential to find a way to combine the high TMR ratio of CoFeB/MgO/CoFeB MTJs and the current-driven switching capability. It was usually the case that lowering the device resistance required for current-driven switching resulted in reduction of TMR ratio and the current density required to observe the switching is of the order of 6 x $10^6$ A/cm$^2$ or higher.[19-23] In this letter, we report current-driven magnetization switching with $J_c$ on the order of $10^5$ A/cm$^2$ in low-resistance Co$_{40}$Fe$_{40}$B$_{20}$/MgO/Co$_{40}$Fe$_{40}$B$_{20}$ MTJs with high TMR ratios.

A schematic diagram of a fabricated MTJ pillar is shown in Fig. 1 (a). MTJ films with synthetic pin layers were deposited on SiO$_2$/Si substrates using RF magnetron sputtering with a base pressure of $10^{-9}$ Torr. The order of the film layers was, starting from the substrate side, Ta(5) / Au(50) / Ta(5) / NiFe(5) / MnIr(8) / CoFe(4) / Ru(0.8) / Co$_{40}$Fe$_{40}$B$_{20}$(6) / MgO (0.85) / Co$_{40}$Fe$_{40}$B$_{20}$($t$) / Ta(5) / Ru(5) (in nm). The thickness $t$ of CoFeB free layer was varied from 1 nm to 3 nm. To reduce the resistance and minimize the chances of electrical breakdown, the thickness of MgO was set to 0.85-nm. The completed MTJs were annealed at annealing temperature ($T_a$) ranging from 270 to 350°C for 1 h in vacuum of $10^{-6}$ Torr under a magnetic field of 4 kOe. All nano-scaled junctions were fabricated using an electron-beam lithography process. A Scanning electron microscopy (SEM) image of a typical MTJ pillar in Fig. 1 (b) shows that the pillar is rectangular in shape with dimensions, 80 x 240 nm$^2$. The TMR loops and current–voltage (*I-V*) characteristics of the fabricated MTJs were measured at room temperature using a four-probe method with a dc bias and a magnetic field of up to 1 kOe. The current direction is defined positive when the electrons are flowing from the top (free) to the bottom (pin) layer as shown in Fig 1 (a).

Figure 2 plots the TMR ratio as a function of $T_a$ for CoFeB/MgO/CoFeB MTJs with a 2-nm

CoFeB free layer and a 0.85-nm MgO barrier; samples were annealed at each $T_a$ successively. As seen in Fig. 2, the TMR ratio increases with increasing $T_a$ and reaches 160 % with corresponding resistance area product (*RA*) of 9 $\Omega\mu m^2$ at $T_a$ = 350°C. These TMR ratios are much greater than those of conventional MTJs using an aluminum oxide barrier having similar *RA*.[20-23] The inset to Fig. 2 shows the bias voltage (left axis) and resistance (right axis) as a function of the applied current for our typical CoFeB/MgO/CoFeB MTJ annealed at $T_a$ = 350°C. The solid and dashed lines show data for parallel and anti-parallel configurations, respectively. Despite the fact that the oxide thickness is thin (0.85 nm) and low *RA*, we noticed no electrical breakdown in the measured bias range of 1 V. In the parallel configuration, the resistance *R* is almost constant, whereas in the anti-parallel configuration *R* decreases as the bias current increases. These high TMR ratio and peculiar *I-V* characteristic are in line with what we reported earlier highly (100)-oriented CoFeB/MgO/CoFeB MTJs,[14] suggesting that the tunneling processes are due to coherent highly-polarized tunneling.

Figures 3 show *RA* versus the current density *J* characteristics at room temperature for samples annealed at 270°C (sample A, Fig. 3 (a)) and annealed at 300°C (sample B, Fig. 3 (b)), with corresponding *RA* versus magnetic-field (*H*) loops measured at a bias voltage of 10 mV as an inset in each panel. The *RA - J* curves were measured under an applied magnetic field of -19 – -20 Oe along the direction of the pin CoFeB layer, in order to compensate the offset (see the two insets) arising from the stray fields of the edge of the patterned synthetic antiferromagnetic pin layer, and in addition perhaps, from a ferromagnetic Neel coupling between the two ferromagnetic layers separated by a very thin MgO barrier. These two samples yielded $RA$ of 7.5 $\Omega\mu m^2$ and the TMR ratio of 49% (sample A) and 11 $\Omega\mu m^2$ and 73% (sample B). The $RA$-$J$ loops showed current-driven sharp resistance switching between the parallel and anti-parallel magnetization alignments, with each MTJ exhibiting a TMR ratio identical to that obtained from the $RA$ - $H$ measurements, indicating full magnetization reversal of the CoFeB free layer by the applied current. The critical current density $J_c$ required to switch the magnetization from parallel (anti-parallel) to anti-parallel (parallel) configuration were +7.8 x $10^5$ A/cm$^2$ (-7.8 x $10^5$ A/cm$^2$) for sample A, and +1.1 x $10^6$ A/cm$^2$ (-6.5 x $10^5$ A/cm$^2$) for sample B. The averaged critical current densities $J_c$ (($J_c^{P->AP}$-$J_c^{AP->P}$)/2) were 7.8 x $10^5$ A/cm$^2$ (sample A) and 8.8 x $10^5$ A/cm$^2$ (sample B). These $J_c$ values are lower than those obtained from MTJs with an aluminum oxide barrier and a single CoFe or CoFeB free-layer structure reported to date.[20-23]

Figure 4 plots $J_c$ as a function of $T_a$ for MTJs with 2 nm-CoFeB free layer (closed circles). The $J_c$ of MTJ with 3 nm-CoFeB free layer (open circle) was also included. $J_c$ increases with increasing $T_a$, while MTJs annealed at higher $T_a$ exhibits a higher TMR ratio. This, we believe, is related to the fact that as-deposited CoFeB film has an amorphous structure that crystallizes when annealed at around 325°C[14,15]. The crystallization of the CoFeB free layer leads to an increase in the saturated magnetization $M_s$; the separately measured $M_s$ of amorphous and crystalline CoFeB layer were 1.3 T and 1.6 T, respectively. We also note that $J_c$ increases with increase of the CoFeB free layer thickness as seen in Fig. 4 at $T_a = 350^{\circ}C$; we see an increase of $J_c$ from 2.5 x $10^6 A/cm^2$ (2nm free layer) to 1.2 x $10^7 A/cm^2$ (3nm free layer).

In the following, we compare our experimental results with the Slonczewski model,[1] because this model qualitatively explains our observation of low $J_c$ for small $M_s$ and small free layer volume. We calculate $J_c$ for sample B annealed at 300°C (Fig. 3 (b)) using the following Slonczewski formula taking into account thermally activated nature of the magnetization reversal,[1,24,25]

$$J_c = J_{c0}\{1-(k_B T/E)\ln(\tau_m f_0)\}, \quad (1)$$

$$J_{c0} = \alpha \gamma e M_s t (H_{ext} \pm H_{ani} \pm H_d)/\mu_B g, \quad (2)$$

$$E = M_s V H_c/2, \quad (3)$$

$$g = [-4+(P^{-1/2}+P^{1/2})(3+\cos\theta)/4]^{-1} \quad (4)$$

where $k_B$ is the Boltzmann constant, $T$ the ambient temperature, $\tau_m$ the measurement time, $f_0$ the attempt frequency, $\alpha$ the Gilbert damping coefficient, $\gamma$ the gyromagnetic constant, $e$ the elementary charge, $t$ the thickness of the free layer, $H_{ext}$ the external magnetic field, $H_{ani}$ the in-plane uniaxial magnetic anisotropy, and $H_d$ the out-of-plane magnetic anisotropy induced by the demagnetization field. $\theta$ is 0 for parallel configuration and $\pi$ for anti-parallel. We assume that the CoFeB free layer annealed at 300°C retains its amorphous structure[14,15]. We used the following parameters to calculate $J_c$: $\alpha = 0.01$, $M_s = 1.3$ T, $t = 2$ nm, $H_c = 1$ Oe, $H_d = 1.3$ T which is equal to $4\pi M_s$, $P = 0.52$ determined from TMR ratio of 73% using Julliere's formula, $\tau_m = 1$ s, $f_0 = 10^9$ Hz. The calculated $J_c$ at 300 K is 4 x $10^6$ A/cm$^2$, a low value owing to the high spin polarization of the tunneling process, but still greater than the experimental value (8.8 x $10^5$ A/cm$^2$).

One of the possible factors contributing to the discrepancy between the calculated and experimental $J_c$ is the effective thickness (volume) of the CoFeB free layer; a magnetic dead layer can be present at the interface of CoFeB that is in contact with Ta cap.[26] Figure 5 plots the TMR ratio as a function of the thickness of CoFeB free layer for MTJs annealed at 300°C. Here, the TMR ratio rapidly decreases with decreasing the thickness of the CoFeB free layer and disappears below 1 nm, which suggests the presence of a magnetic dead layer of the order of 1 nm in the CoFeB free layer. This justifies the use of $t$ reduced from the nominal thickness: for example, if we take $t = 1.5$ nm (a 0.5 nm thick dead layer), we obtain $J_c$ of 8 x $10^5$ A/cm$^2$, in reasonably good agreement with the experimentally observed $J_c$. We can thus attribute the low $J_c$ on the order of $10^5$ A/cm$^2$ to high spin-polarization and the small $M_sV$ product of the CoFeB single free layer. We note that the thermal stability factor ($K_uV/kT$, where $K_u$ is the uniaxial anisotropy constant), which needs to be greater than 60 to ensure the stability over 10 years, is not fulfilled in our device. The stability of the free layer, however, can be enhanced for example by adding a synthetic antiferromagnet to the free layer.[27,28]

In conclusion, we have described current-driven magnetization switching in low-resistance $Co_{40}Fe_{40}B_{20}$/MgO/$Co_{40}Fe_{40}B_{20}$ magnetic tunnel junctions (MTJs) with high TMR ratio. In MTJs annealed at 270$^{o}$C, the magnetization switching occurred at a critical current density of 7.8 x 10$^5$ A/cm$^2$. Annealing the samples at 350$^{o}$C increased the TMR ratio to 160 % but $J_c$ also increased to 2.5 x 10$^6$ A/cm$^2$. The observed $J_c$ on the order of 10$^5$ A/cm$^2$ is attributed to high spin-polarized tunnel current and the small $M_sV$ of the initially amorphous CoFeB single free layer. This MTJ structure with $J_c$ on the order of 10$^5$ A/cm$^2$ and high TMR ratio has potential for the future MRAM.

This work was supported by the IT-program of Research Revolution 2002 (RR2002): "Development of Universal Low-power Spin Memory", Ministry of Education, Culture, Sports, Science and Technology of Japan.

A. Buhrman: Appl. Phys. Lett. 86 (2005) 152509.

24) R. H. Koch, J. A. Katine, and J. Z. Sun: Phys. Rev. Lett. 92 (2004) 088302.

25) D. Lacour, J. A. Katine, N. Smith, M. J. Carey, and J. R. Childress: Appl. Phys. Lett. 85 (2004) 4681.

26) C.-Y. Hung, M. Mao, S. Funada, T. Schneider, L. Miloslavsky, M. Miller, C. Qian, and H. C. Tong: J. Appl. Phys. 87 (2000) 6618.

27) J. Janesky, N. D. Rizzo, B. N. Engel, and S. Tehrani: Appl. Phys. Lett. 85 (2004) 2289.

28) T. Ochiai, Y. Jiang, A. Hirohata, N. Tezuka, S. Sugimoto, and K. Inomata: Appl. Phys. Lett. 86 (2005) 242506.
11

Figure captions

Fig. 1. (a) Schematic drawing of the cross section of fabricated MTJ. The thickness of MgO is fixed to 0.85 nm. (b) Scanning electron microscopy image of a pillar after milling showing that the area of the fabricated pillar is 80 x 240 nm$^2$.

Fig. 2. TMR ratios as function of annealing temperature ($T_a$) for CoFeB/MgO/CoFeB MTJs with a 2-nm CoFeB. The inset shows bias voltage (left axis) and resistance (right axis) as a function of current for an MTJ annealed at $T_a$ = 350$^o$C. Solid and dashed lines show data in parallel and anti-parallel configurations, respectively.

Fig.3. Resistance area product (*RA*) versus current density (*J*) loops at room temperature for two samples annealed (a) at 270$^o$C (sample A) and (b) at 300$^o$C (sample B). These curves were taken under a fixed external magnetic field to compensate the effect of the offset field. The insets show *RA* versus magnetic-field (*H*) loops. The averaged critical current density $J_c$ (($J_c^{P->AP}$-$J_c^{AP->P}$)/2) is 7.8 x 10$^5$ A/cm$^2$ (with TMR ratio of 49%) for sample A and 8.8 x 10$^5$ A/cm$^2$ (with TMR ratio of 73%) for sample B.

Fig.4. $J_c$ as a function of $T_a$ for MTJs with 2- (closed circle) and 3-nm (open circle) CoFeB free

layer.

Fig.5. The CoFeB free layer thickness dependence of TMR ratios for CoFeB/MgO/CoFeB MTJs annealed at 300°C. TMR ratio decreases with decreasing the free layer thickness and goes to zero when the thickness is below 1 nm.

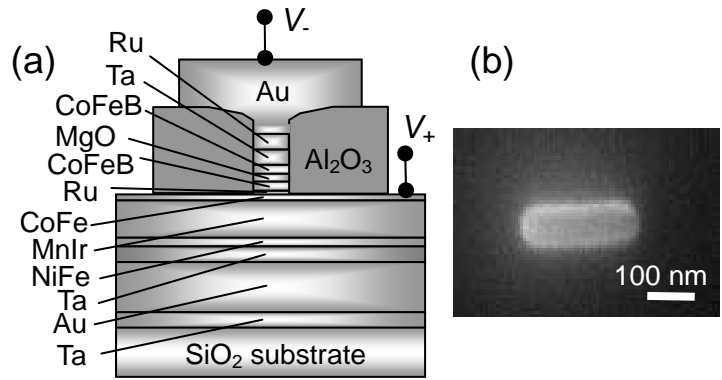

Fig. 1   J. Hayakawa et al.

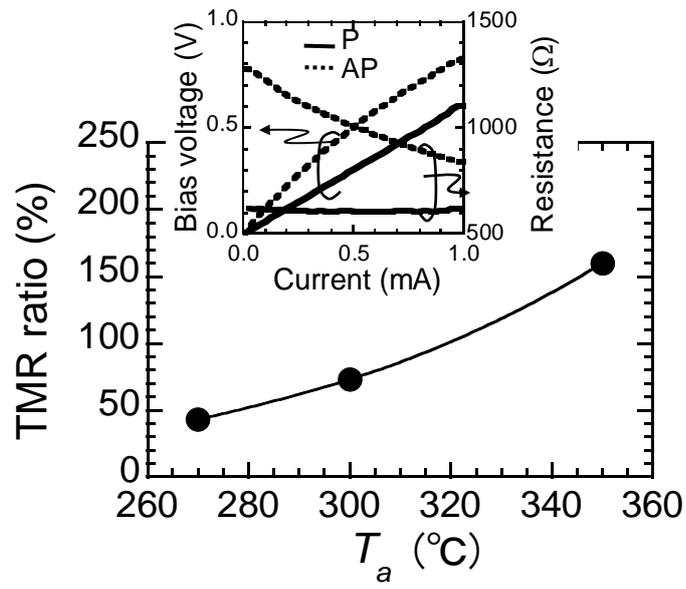

Fig. 2 J. Hayakawa et al.

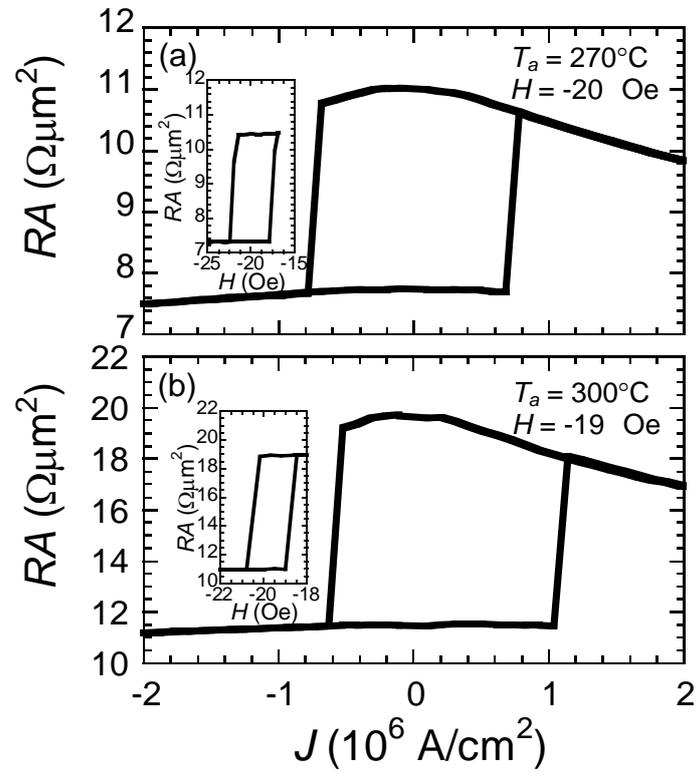

Fig. 3 J. Hayakawa et al.

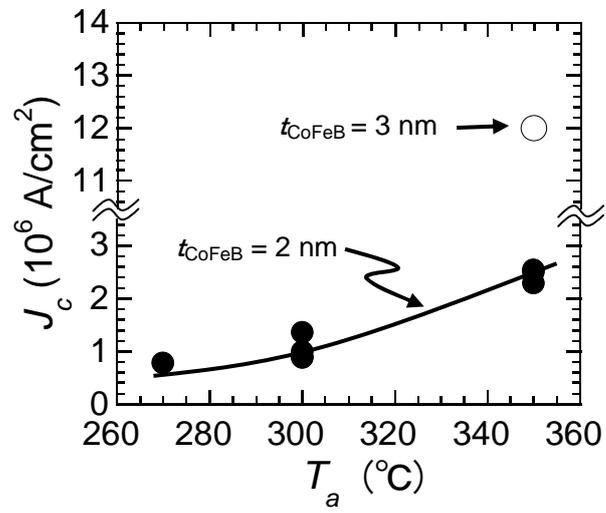

Fig. 4 J. Hayakawa et al.

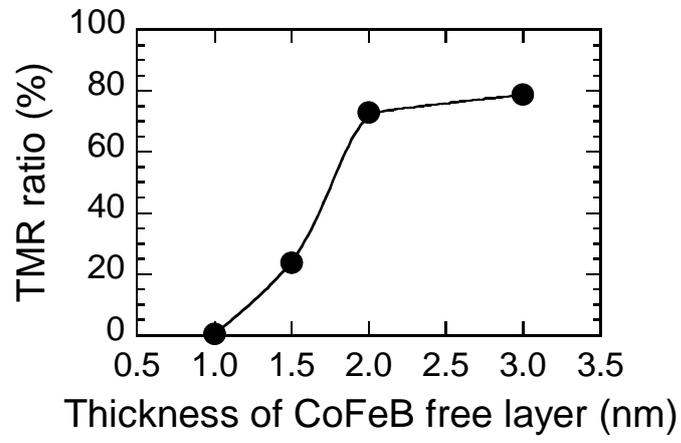

Fig. 5 J. Hayakawa et al.